\documentclass[11pt]{article}
\usepackage{amsmath,amssymb,amsfonts,bbm}

\newcommand{\be}{\begin{equation}}
\newcommand{\ee}{\end{equation}}
\newcommand{\bea}{\begin{eqnarray}}
\newcommand{\eea}{\end{eqnarray}}
\newcommand{\ba}{\begin{array}}
\newcommand{\ea}{\end{array}}

\newcommand{\htwo}{h_{2,1}}
\newcommand{\M}{\mathcal{M}}
\newcommand{\N}{\mathcal{N}}

\newcommand{\K}{\mathcal{K}}

\long\def\symbolfootnote[#1]#2{\begingroup%
\def\thefootnote{\fnsymbol{footnote}}\footnote[#1]{#2}\endgroup}

\begin{document}

\thispagestyle{empty}\vspace{40pt}

\hfill{}

\vspace{128pt}

\begin{center}
    \textbf{\Large BPS one-branes in five dimensions}\\
    \vspace{40pt}

    Moataz H. Emam\symbolfootnote[1]{\tt moataz.emam@cortland.edu}

    \vspace{12pt}   \textit{Department of Physics}\\
                    \textit{SUNY College at Cortland}\\
                    \textit{Cortland, NY 13045, USA}\\
\end{center}

\vspace{40pt}

\begin{abstract}

In a recent paper, we studied the scalar fields of the five dimensional $\N=2$ hypermultiplets using the method of symplectic covariance. For static spherically symmetric backgrounds, we showed that exactly two possibilities exist and detailed one of them. In this paper, we present the second case: Bogomol'nyi-Prasad-Sommerfield (BPS) one-branes carrying charge under the hypermultiplets. We find an explicitly symplectic solution of the fields in this background and derive the conditions required for a full spacetime understanding.

\end{abstract}

\newpage

%%%%%%%%%%%%%%%%%%%%%%%%%%%%%%%%%%%%%%%%%%%%%%%%%%%%%%%%%%%%%%%%%%%%%%%%%%

%\tableofcontents

\vspace{15pt}

\pagebreak

\section{Introduction}

Studies of $\N=2$ supergravity theories in four as well as five dimensions usually focus on the vector and/or tensor multiplet regimes. This is due to the fact that their underlying special K\"{a}hler geometry is very well understood (see for example \cite{Butter:2012xg,Klemm:2009uw,Mohaupt:2008zz,Cacciatori:2008ek,Cortes:2003zd,Isozumi:2003uh,Cacciatori:2003kv,Andrianopoli:2011zj} and references within). On the other hand, solutions in the hypermultiplets sector are rare, due in part to the mathematical complexity involved, since the hypermultiplets generally parameterize quaternionic manifolds \cite{de Wit:2001dj}. However, it was pointed out some years ago that due to the so-called $c$-map, the hypermultiplets in $D=5$ for instance can be related to the much better understood $D=4$ vector multiplets, and that the methods of special geometry, developed for the latter, can be applied to the former \cite{Gutperle:2000ve}. Based on this observation, some hypermultiplet constructions in instanton and certain two-brane backgrounds were found and studied (last reference and \cite{Emam:2005bh,Emam:2006sr}). More recently, we argued in \cite{Emam:2009xj} that the well known symplectic structure of quaternionic and special  K\"{a}hler manifolds \cite{deWit:1995jd} can be used to construct hypermultiplet ``solutions'' based on covariance in symplectic space. These are full solutions \emph{only} in the symplectic sense, written in terms of symplectic basis vectors and invariants. In \cite{Emam:2012zw} we applied the symplectic method in an effort to classify general hypermultiplet solutions in backgrounds of maximal symmetry (spherically symmetric backgrounds etc), and showed that exactly two cases can arise. The first case, studied in the aforementioned paper, is that of BPS zero-branes. In this sequel, we present the second case: $D=5$ BPS one-branes coupled to the full set of hypermultiplet fields. We find some explicit spacetime solutions to the hyperscalars and derive constraints on the complex structure moduli of the underlying Calabi-Yau that may be used in future work for a deeper understanding.

\section{$D=5$ $\N=2$ supergravity with hypermultiplets} \label{theory}

The dimensional reduction of $D=11$ supergravity theory over a Calabi-Yau 3-fold $\M$ with nontrivial complex structure moduli yields an $\N=2$ supergravity theory in $D=5$ with a set of scalar fields and their supersymmetric partners all together known as the \emph{hypermultiplets} (See \cite{Emam:2010kt} for a review and additional references). These are partially comprised of the \emph{universal hypermultiplet} $\left(a, \sigma, \zeta^0, \tilde \zeta_0\right)$, so called because it appears irrespective of the detailed structure of the Calabi-Yau. The field $a$ is known as the universal axion and the dilaton $\sigma$ is proportional to the natural logarithm of the volume of $\M$. The rest of the hypermultiplets are $\left(z^i, z^{\bar i}, \zeta^i, \tilde \zeta_i: i=1,\ldots, \htwo\right)$, where the $z$'s are identified with the complex structure moduli of $\M$, and $\htwo$ is the Hodge number determining the dimensions of the manifold of the Calabi-Yau's complex structure moduli; $\M_C$. The `bar' over an index denotes complex conjugation. The fields $\left(\zeta^I, \tilde\zeta_I: I=0,\ldots,\htwo\right)$ are known as the axions and arise as a result of the $D=11$ Chern-Simons term. The supersymmetric partners known as the hyperini complete the hypermultiplets. The axionic fields $\left(\zeta^I, \tilde\zeta_I\right)$ can be defined as components of a symplectic vector $\left| \Xi  \right\rangle$ such that the symplectic scalar product is defined by, for example:
\be
    \left\langle {{\Xi }}
 \mathrel{\left | {\vphantom {{\Xi } d\Xi }}
 \right. \kern-\nulldelimiterspace}
 {d\Xi } \right\rangle   = \zeta^I d\tilde \zeta_I  - \tilde \zeta_I
 d\zeta^I,\label{DefOfSympScalarProduct}
\ee
where $d$ is the spacetime exterior derivative $\left(d=dx^M\partial_M:M=0,\ldots,4\right)$. One can define the symplectic basis vectors $\left| V \right\rangle $, $\left| {U_i } \right\rangle $ and their complex conjugates such that
\bea
    \left\langle {{\bar V}}
     \mathrel{\left | {\vphantom {{\bar V} V}}
     \right. \kern-\nulldelimiterspace}
     {V} \right\rangle   &=& i\nonumber\\
    \left|\nabla _i  {\bar V} \right\rangle  &=& \left|\nabla _{\bar i}  V \right\rangle =0\nonumber\\
    \left\langle {{U_i }}
    \mathrel{\left | {\vphantom {{U_i } {U_j }}}
    \right. \kern-\nulldelimiterspace}
    {{U_j }} \right\rangle  &=& \left\langle {{U_{\bar i} }}
    \mathrel{\left | {\vphantom {{U_{\bar i} } {U_{\bar j} }}}
    \right. \kern-\nulldelimiterspace}
    {{U_{\bar j} }} \right\rangle    =
    \left\langle {\bar V}
    \mathrel{\left | {\vphantom {\bar V {U_i }}}
    \right. \kern-\nulldelimiterspace}
    {{U_i }} \right\rangle  = \left\langle {V}
    \mathrel{\left | {\vphantom {V {U_{\bar i} }}}
    \right. \kern-\nulldelimiterspace}
    {{U_{\bar i} }} \right\rangle  = \left\langle { V}
    \mathrel{\left | {\vphantom { V {U_i }}}
    \right. \kern-\nulldelimiterspace}
    {{U_i }} \right\rangle=\left\langle {\bar V}
    \mathrel{\left | {\vphantom {\bar V {U_{\bar i} }}}
    \right. \kern-\nulldelimiterspace}
    {{U_{\bar i} }} \right\rangle= 0,\nonumber\\
    \left|\nabla _{\bar j}  {U_i } \right\rangle  &=& G_{i\bar j} \left| V \right\rangle ,\quad \quad \left|\nabla _i  {U_{\bar j} } \right\rangle  = G_{i\bar j} \left| {\bar V}
    \right\rangle,\quad\quad
    G_{i\bar j}= \left( {\partial _i \partial _{\bar j} \K} \right)=- i    \left\langle {{U_i }}
    \mathrel{\left | {\vphantom {{U_i } {U_{\bar j} }}}
    \right. \kern-\nulldelimiterspace}
    {{U_{\bar j} }} \right\rangle,
\eea
where the covariant derivatives are with respect to the moduli $\left(z^i, z^{\bar i}\right)$ and $G_{i\bar j}$ is a special K\"{a}hler metric on $\M_C$. The origin of these identities lies in special K\"{a}hler geometry. In our paper \cite{Emam:2009xj}, we derived the following useful formulae:
\bea
    dG_{i\bar j}  &=& G_{k\bar j} \Gamma _{ri}^k dz^r  + G_{i\bar k} \Gamma _{\bar r\bar j}^{\bar k} dz^{\bar r}  \nonumber\\
    dG^{i\bar j}  &=&  - G^{p\bar j} \Gamma _{rp}^i dz^r  - G^{i\bar p} \Gamma _{\bar r\bar p}^{\bar j} dz^{\bar r}  \nonumber\\
    \left| {dV} \right\rangle  &=& dz^i \left| {U_i } \right\rangle  - i\mathfrak{Im} \left[ {\left( {\partial_i  \K} \right)dz^i} \right]\left| V \right\rangle \nonumber \\
    \left| {d\bar V} \right\rangle  &=& dz^{\bar i} \left| {U_{\bar i} } \right\rangle  + i\mathfrak{Im} \left[ {\left( {\partial_i  \K} \right)dz^i} \right]\left| {\bar V} \right\rangle \nonumber \\
    \left| {dU_i } \right\rangle  &=& G_{i\bar j} dz^{\bar j} \left| V \right\rangle  + \Gamma _{ik}^r dz^k \left| {U_r } \right\rangle+G^{j\bar l} C_{ijk} dz^k \left| {U_{\bar l} } \right\rangle - i\mathfrak{Im} \left[ {\left( {\partial_i  \K} \right)dz^i} \right]\left| {U_i } \right\rangle \nonumber \\
    \left| {dU_{\bar i} } \right\rangle  &=& G_{j\bar i} dz^j \left| {\bar V} \right\rangle + \Gamma _{\bar i\bar k}^{\bar r} dz^{\bar k} \left| {U_{\bar r} } \right\rangle + G^{l\bar j} C_{\bar i\bar j\bar k} dz^{\bar k} \left| {U_l } \right\rangle + i\mathfrak{Im} \left[ {\left( {\partial_i  \K} \right)dz^i} \right]\left| {U_{\bar i} } \right\rangle\nonumber\\
     {\bf \Lambda } &=& 2\left| V \right\rangle \left\langle {\bar V} \right| + 2G^{i\bar j} \left| {U_{\bar j} } \right\rangle \left\langle {U_i } \right|
    -i\nonumber\\
    {\bf \Lambda }^{-1} &=& -2\left| V \right\rangle \left\langle {\bar V} \right| - 2G^{i\bar j} \left| {U_{\bar j} } \right\rangle \left\langle {U_i } \right|
    +i\nonumber\\
        \partial_i {\bf \Lambda } &=& 2\left| {U_i } \right\rangle \left\langle {\bar V} \right|+2\left| {\bar V} \right\rangle \left\langle {U_i } \right| + 2G^{j\bar r} G^{k\bar p} C_{ijk} \left| {U_{\bar r} } \right\rangle \left\langle {U_{\bar p} } \right|.\label{Myequations}
\eea
where $\K$ is the K\"{a}hler potential of $\M_C$ and the quantities $C_{ijk}$ are the components of the totally symmetric tensor that appears in the curvature tensor of $\M_C$. The symplectic matrix $\bf \Lambda$ is viewed as a rotation matrix in symplectic space. In this language, the bosonic part of the action is:
\bea
    S_5  &=& \int\limits_5 {\left[ {R\star \mathbf{1} - \frac{1}{2}d\sigma \wedge\star d\sigma  - G_{i\bar j} dz^i \wedge\star dz^{\bar j} } \right.}  + e^\sigma   \left\langle {d\Xi } \right|\mathop{\bf \Lambda} \limits_ \wedge  \left| {\star d\Xi } \right\rangle\nonumber\\
    & &\left. {\quad\quad\quad\quad\quad\quad\quad\quad\quad\quad\quad\quad\quad - \frac{1}{2} e^{2\sigma } \left[ {da + \left\langle {\Xi } \mathrel{\left | {\vphantom {\Xi  {d\Xi }}} \right. \kern-\nulldelimiterspace} {{d\Xi }}    \right\rangle} \right] \wedge \star\left[ {da + \left\langle {\Xi } \mathrel{\left | {\vphantom {\Xi  {d\Xi }}} \right. \kern-\nulldelimiterspace} {{d\Xi }}    \right\rangle} \right] } \right],\label{action}
\eea
where $\star$ is the $D=5$ Hodge duality operator. The variation of the action yields the following field equations for $\sigma$, $\left(z^i,z^{\bar i}\right)$, $\left| \Xi  \right\rangle$ and $a$ respectively:
\bea
    \left( {\Delta \sigma } \right)\star \mathbf{1} + e^\sigma   \left\langle {d\Xi } \right|\mathop {\bf \Lambda} \limits_ \wedge  \left| {\star d\Xi } \right\rangle -   e^{2\sigma }\left[ {da + \left\langle {\Xi } \mathrel{\left | {\vphantom {\Xi  {d\Xi }}} \right. \kern-\nulldelimiterspace} {{d\Xi }}    \right\rangle} \right]\wedge\star\left[ {da + \left\langle {\Xi } \mathrel{\left | {\vphantom {\Xi  {d\Xi }}} \right. \kern-\nulldelimiterspace} {{d\Xi }}    \right\rangle} \right] &=& 0\label{DilatonEOM}\\
    \left( {\Delta z^i } \right)\star \mathbf{1} + \Gamma _{jk}^i dz^j  \wedge \star dz^k  + \frac{1}{2}e^\sigma  G^{i\bar j}  {\partial _{\bar j} \left\langle {d\Xi } \right|\mathop {\bf \Lambda} \limits_ \wedge  \left| {\star d\Xi } \right\rangle} &=& 0 \nonumber\\
    \left( {\Delta z^{\bar i} } \right)\star \mathbf{1} + \Gamma _{\bar j\bar k}^{\bar i} dz^{\bar j}  \wedge \star dz^{\bar k}  + \frac{1}{2}e^\sigma  G^{\bar ij}  {\partial _j \left\langle {d\Xi } \right|\mathop {\bf \Lambda} \limits_ \wedge  \left| {\star d\Xi } \right\rangle}  &=& 0\label{ZZBarEOM} \\
    d^{\dag} \left\{ {e^\sigma  \left| {{\bf \Lambda} d\Xi } \right\rangle  - e^{2\sigma } \left[ {da + \left\langle {\Xi }
    \mathrel{\left | {\vphantom {\Xi  {d\Xi }}}\right. \kern-\nulldelimiterspace} {{d\Xi }} \right\rangle } \right]\left| \Xi  \right\rangle } \right\} &=& 0\label{AxionsEOM}\\
    d^{\dag} \left[ {e^{2\sigma } da + e^{2\sigma } \left\langle {\Xi } \mathrel{\left | {\vphantom {\Xi  {d\Xi }}} \right. \kern-\nulldelimiterspace} {{d\Xi }}    \right\rangle} \right] &=&    0\label{aEOM}
\eea
where $d^\dagger$ is the $D=5$ adjoint exterior derivative, $\Delta$ is the Laplace-de Rahm operator and $\Gamma _{jk}^i$ is the connection on $\M_C$. The full action is symmetric under the following SUSY transformations:
\bea
 \delta _\epsilon  \psi ^1  &=& D \epsilon _1  + \frac{1}{4}\left\{ {i {e^{\sigma } \left[ {da + \left\langle {\Xi }
 \mathrel{\left | {\vphantom {\Xi  {d\Xi }}}
 \right. \kern-\nulldelimiterspace} {{d\Xi }} \right\rangle } \right]}- Y} \right\}\epsilon _1  - e^{\frac{\sigma }{2}} \left\langle {{\bar V}}
 \mathrel{\left | {\vphantom {{\bar V} {d\Xi }}} \right. \kern-\nulldelimiterspace} {{d\Xi }} \right\rangle\epsilon _2  \nonumber\\
 \delta _\epsilon  \psi ^2  &=& D \epsilon _2  - \frac{1}{4}\left\{ {i {e^{\sigma } \left[ {da + \left\langle {\Xi }
 \mathrel{\left | {\vphantom {\Xi  {d\Xi }}} \right. \kern-\nulldelimiterspace}
 {{d\Xi }} \right\rangle } \right]}- Y} \right\}\epsilon _2  + e^{\frac{\sigma }{2}} \left\langle {V}
 \mathrel{\left | {\vphantom {V {d\Xi }}} \right. \kern-\nulldelimiterspace} {{d\Xi }} \right\rangle \epsilon _1 \\ \label{SUSYGraviton}
  \delta _\epsilon  \xi _1^0  &=& e^{\frac{\sigma }{2}} \left\langle {V}
    \mathrel{\left | {\vphantom {V {\partial _M  \Xi }}} \right. \kern-\nulldelimiterspace} {{\partial _M  \Xi }} \right\rangle  \Gamma ^M  \epsilon _1  - \left\{ {\frac{1}{2}\left( {\partial _M  \sigma } \right) - \frac{i}{2} e^{\sigma } \left[ {\left(\partial _M a\right) + \left\langle {\Xi }
    \mathrel{\left | {\vphantom {\Xi  {\partial _M \Xi }}} \right. \kern-\nulldelimiterspace}
    {{\partial _M \Xi }} \right\rangle } \right]} \right\}\Gamma ^M  \epsilon _2  \nonumber\\
     \delta _\epsilon  \xi _2^0  &=& e^{\frac{\sigma }{2}} \left\langle {{\bar V}}
    \mathrel{\left | {\vphantom {{\bar V} {\partial _M  \Xi }}} \right. \kern-\nulldelimiterspace} {{\partial _M  \Xi }} \right\rangle \Gamma ^M  \epsilon _2  + \left\{ {\frac{1}{2}\left( {\partial _M  \sigma } \right) + \frac{i}{2} e^{\sigma } \left[ {\left(\partial _M a\right) + \left\langle {\Xi }
    \mathrel{\left | {\vphantom {\Xi  {\partial _M \Xi }}} \right. \kern-\nulldelimiterspace}
    {{\partial _M \Xi }} \right\rangle } \right]} \right\}\Gamma ^M  \epsilon
     _1\label{SUSYHyperon1}\\
     \delta _\epsilon  \xi _1^{\hat i}  &=& e^{\frac{\sigma }{2}} e^{\hat ij} \left\langle {{U_j }}
    \mathrel{\left | {\vphantom {{U_j } {\partial _M  \Xi }}} \right. \kern-\nulldelimiterspace} {{\partial _M  \Xi }} \right\rangle \Gamma ^M  \epsilon _1  - e_{\,\,\,\bar j}^{\hat i} \left( {\partial _M  z^{\bar j} } \right)\Gamma ^M  \epsilon _2  \nonumber\\
     \delta _\epsilon  \xi _2^{\hat i}  &=& e^{\frac{\sigma }{2}} e^{\hat i\bar j} \left\langle {{U_{\bar j} }}
    \mathrel{\left | {\vphantom {{U_{\bar j} } {\partial _M  \Xi }}} \right. \kern-\nulldelimiterspace} {{\partial _M  \Xi }} \right\rangle \Gamma ^M  \epsilon _2  + e_{\,\,\,j}^{\hat i} \left( {\partial _M  z^j } \right)\Gamma ^M  \epsilon    _1,\label{SUSYHyperon2}
\eea
where $\left(\psi ^1, \psi ^2\right)$ are the two gravitini and $\left(\xi _1^I, \xi _2^I\right)$ are the hyperini. The quantity $Y$ is related to the periods $\left(Z^I,F_I\right)$ of the Calabi-Yau's unique holomorphic form as follows:
\begin{equation}
    Y   = 2ie^{-\K}\mathfrak{Im}\left[  \left({\partial_IF_J } \right) \bar Z^I {d  Z^J }  \right].\label{DefOfY}
\end{equation}

The $e_{\,\,\,j}^{\hat i}$'s are the beins of the special K\"{a}hler metric $G_{i\bar j}$, the $\epsilon$'s are the five dimensional $\N=2$ SUSY spinors and the $\Gamma^M$'s are the usual Dirac matrices. Finally, the covariant derivative $D$ is
\be
    D=dx^M\left( \partial _M   + \frac{1}{4}\omega _M^{\,\,\,\,\hat M\hat N} \Gamma _{\hat M\hat
    N}\right)\label{DefOfCovDerivative}
\ee
as usual, where the $\omega$'s are the spin connections and the hatted indices are frame indices in a flat tangent space.

\section{Analysis and results}\label{Brane Analysis}

The most general spherically symmetric $p$-branes in $D=5$ can be represented by the following (Poincar\'{e})$_{p+1} \times SO\left(4-p\right)$ metric:
\be
    ds^2  = e^{2C\sigma } \eta _{ab} dx^a dx^b  + e^{2B\sigma } \delta _{\mu \nu } dx^\mu  dx^\nu,\label{GeneralMetric}
\ee
where $B$ and $C$ are constants, the directions $a,b=0,1,\ldots,p$ define the brane's world-volume while $\mu,\nu=(p+1),\ldots,4$ are those transverse to the brane. The dilaton is assumed purely radial in the $\mu,\nu$ directions. It turns out that the constant $C$ is constrained to vanish by both the Einstein equations and the SUSY condition\footnote{This is no surprise, since $\delta \psi=0$ automatically satisfies $G_{MN}= T_{MN}$ \cite{Kaya:1999mm}.} $\delta \psi=0$. We then set $C=0$ from the start and are left only with the task of specifying $B$ and the allowed values of $p$.
The universal axion's field equation (\ref{aEOM}) implies a solution of the form
\be
     da +\left\langle {\Xi }
 \mathrel{\left | {\vphantom {\Xi  {d\Xi }}}
 \right. \kern-\nulldelimiterspace}
 {{d\Xi }} \right\rangle = \alpha  e^{-2\sigma } dH,
\ee
where $H$ is an arbitrary function satisfying $\Delta H = 0$, and $\alpha  \in \mathbb{R}$. Similarly, the axions field equation (\ref{AxionsEOM}) leads to
\be
    e^\sigma  \left| {{\bf \Lambda} d\Xi } \right\rangle  - \alpha dH\left| \Xi  \right\rangle  = \beta \left| {dK} \right\rangle \,\,\,{\rm where}\,\,\,\left| {\Delta K} \right\rangle  = 0\,\,\,{\rm and}\,\,\,\beta  \in \mathbb{R}.\label{Axion-K}
\ee

Since we are only interested in Bosonic solutions, we consider the vanishing of the supersymmetric variations (\ref{SUSYHyperon1}, \ref{SUSYHyperon2}). These may be rewritten as matrix equations with vanishing determinants:
\bea
 d\sigma  \wedge \star d\sigma  + \alpha ^2 e^{ - 2\sigma } dH \wedge \star dH + 4e^\sigma  \left\langle {V}
 \mathrel{\left | {\vphantom {V {d\Xi }}}
 \right. \kern-\nulldelimiterspace}
 {{d\Xi }} \right\rangle  \wedge \left\langle {{\bar V}}
 \mathrel{\left | {\vphantom {{\bar V} {\star d\Xi }}}
 \right. \kern-\nulldelimiterspace}
 {{\star d\Xi }} \right\rangle  &=& 0 \nonumber\\
 G_{i\bar j} dz^i  \wedge \star dz^{\bar j}  + e^\sigma  G^{i\bar j} \left\langle {{U_i }}
 \mathrel{\left | {\vphantom {{U_i } {d\Xi }}}
 \right. \kern-\nulldelimiterspace}
 {{d\Xi }} \right\rangle  \wedge \left\langle {{U_{\bar j} }}
 \mathrel{\left | {\vphantom {{U_{\bar j} } {\star d\Xi }}}
 \right. \kern-\nulldelimiterspace}
 {{\star d\Xi }} \right\rangle  &=& 0.\label{FromSYSY}
\eea

Using this with the definition of the symplectic matrix $\bf\Lambda$ in (\ref{Myequations}) we find
\be
    e^\sigma  \left\langle {d\Xi } \right|\mathop {\bf \Lambda} \limits_ \wedge  \left| {\star d\Xi } \right\rangle  = \frac{1}{2}d\sigma  \wedge \star d\sigma  + \frac{1}{2}\alpha ^2 e^{ - 2\sigma } dH \wedge \star dH + 2G_{i\bar j} dz^i  \wedge \star dz^{\bar j},\label{Rotation}
\ee
where we have used $\left\langle {d\Xi } \right.\mathop |\limits_ \wedge  \left. {\star d\Xi } \right\rangle  = 0$ as required by the reality of the axions. The dilaton's equation (\ref{DilatonEOM}) then becomes
\be
    \left( {\Delta \sigma } \right)\star \mathbf{1} + \frac{1}{2}d\sigma  \wedge \star d\sigma  = \frac{1}{2}\alpha ^2 e^{ - 2\sigma } dH \wedge \star dH - 2G_{i\bar j} dz^i  \wedge \star dz^{\bar j}.\label{dilatonian}
\ee

Finally, the components of the Einstein equations reduce to
\bea
 \frac{1}{2}B\left( {3 - p} \right)\left[ {B\left( {2 - p} \right) - 1} \right]d\sigma  \wedge \star d\sigma  + \left[ {\frac{1}{2} - 2B\left( {3 - p} \right)} \right]G_{i\bar j} dz^i  \wedge \star dz^{\bar j}  \nonumber\\
  =  - \frac{1}{2}B\left( {3 - p} \right)\alpha ^2 e^{ - 2\sigma } dH \wedge \star dH \nonumber\\
 \frac{1}{2}B\left( {2 - p} \right)\left( {2B + 1} \right)d\sigma  \wedge \star d\sigma  + \left[ {2B\left( {2 - p} \right) - 1} \right]G_{i\bar j} dz^i  \wedge \star dz^{\bar j}  \nonumber\\
  = \frac{1}{2}B\left( {2 - p} \right)\alpha ^2 e^{ - 2\sigma } dH \wedge \star dH \nonumber\\
 \frac{1}{2}B\left( {2 - p} \right)\left[ {B\left( {1 - p} \right) - 1} \right]d\sigma  \wedge \star d\sigma  + \left[ {\frac{1}{2} - 2B\left( {2 - p} \right)} \right]G_{i\bar j} dz^i  \wedge \star dz^{\bar j}  \nonumber\\
  =  - \frac{1}{2}B\left( {2 - p} \right)\alpha ^2 e^{ - 2\sigma } dH \wedge \star dH.\label{EinsteinEquations}
\eea

It can be easily shown that equations (\ref{EinsteinEquations}) cannot be simultaneously satisfied for the case $\alpha=0$. They either lead to an imaginary $B$ or to trivial solutions with constant complex structure moduli. On the other hand, the case of non-vanishing $\alpha$ leads to exactly two non-trivial solutions. These are $p=0,1$. We have previously studied the zero-brane case \cite{Emam:2012zw}. We now present the case of the one-brane. For $p=1$, equations (\ref{Rotation}), (\ref{dilatonian}) and (\ref{EinsteinEquations}) are identically satisfied for any value of the constant $B$ if:
\bea
 G_{i\bar j} dz^i  \wedge \star dz^{\bar j}  &=& 2B^2 d\sigma  \wedge \star d\sigma  \nonumber\\
 \left( {8B^2  - 2B + 1} \right)d\sigma  \wedge \star d\sigma  &=& \alpha ^2 e^{ - 2\sigma } dH \wedge \star dH \nonumber\\
 e^\sigma  \left\langle {d\Xi } \right|\mathop {\bf \Lambda} \limits_ \wedge  \left| {\star d\Xi } \right\rangle  &=& \left( {8B^2  - B + 1} \right)d\sigma  \wedge \star d\sigma  \nonumber\\
 \Delta e^{B\sigma }  &=& 0.\label{p0solutionconditions}
\eea

The last equation of (\ref{p0solutionconditions}) implies the simple ansatz $e^{B\sigma }  = H$, which leads to $B = 1$ and $\alpha ^2  = 7$. Hence, the dilaton is fully specified in terms of $H$:
\be
 \sigma  = \ln H.
\ee

To find an expression for the axions, we look again at the vanishing of the hyperini transformations (\ref{SUSYHyperon1}) and (\ref{SUSYHyperon2}) and make the simplifying assumption $\epsilon_1=\pm\epsilon_2$. This leads to:
\be
 \left| {d\Xi } \right\rangle  = e^{ - \frac{\sigma }{2}} \mathfrak{Re} \left[ {\left( {\alpha  - i} \right)\left| V \right\rangle d\sigma  + 2i\left| {U_i } \right\rangle dz^i } \right].\label{dXi}
\ee

One can now substitute (\ref{dXi}) in (\ref{Axion-K}) to get
\be
   \left| \Xi  \right\rangle d\sigma
    = \frac{1}{\alpha }e^{ - \frac{\sigma }{2}} \mathfrak{Re}\left[\left(1+i\alpha\right)\left| V \right\rangle d\sigma\right]
    + \frac{2}{\alpha }e^{ - \frac{\sigma }{2}} \mathfrak{Re}\left[\left| U_i \right\rangle dz^i\right]-\frac{\beta}{\alpha }e^{ - \sigma }\left| {dK} \right\rangle.\label{42}
\ee

Equations (\ref{ZZBarEOM}) are simplified as a consequence of the third result of (\ref{p0solutionconditions}). They reduce to
\be
    \left( {\Delta z^i } \right)\star \mathbf{1} + \Gamma _{jk}^i dz^j  \wedge \star dz^k  = 0,\label{zEOMmodified}
\ee
and similarly for its complex conjugate counterpart. Assuming the ansatz
\be
    dz^i  =  f^i de^{\sigma }\label{MC Flow parameters}
\ee
results in the condition
\be
    df^i  + \Gamma _{jk}^i f^j f^k de^\sigma   = 0.\label{DiffCon}
\ee

Equation (\ref{42}) can now be rewritten
\be
   \left| \Xi  \right\rangle
    = \frac{1}{\alpha }e^{ - \frac{\sigma }{2}} \mathfrak{Re}\left[\left(1+i\alpha\right)\left| V \right\rangle \right] \nonumber\\
    + \frac{2}{\alpha }e^{  \frac{\sigma }{2}} \mathfrak{Re}\left[f^i\left| U_i \right\rangle \right],\label{42again}
\ee
where we have chosen $\beta=0$. The first of equations (\ref{p0solutionconditions}) leads to the additional constraint:
\be
    G_{i\bar j} f^i f^{\bar j}  = 2e^{ - 2\sigma }.
\ee

Using these results, we find
\bea
 da &=& \left( {\frac{{7 - 2\alpha ^2 }}{{2\alpha }}} \right)dH^{ - 1} \,\,\,\,\,{\rm and} \nonumber\\
 a &=& c - \left(\frac{2\alpha^2-7}{2\alpha}\right)\frac{1}{H},
\eea
where $c$ is an arbitrary integration constant related to the asymptotic value of $a$. Finally, solving the SUSY condition $\delta \psi = 0$ gives the following form for the spinors
\bea
    \epsilon _1  &=& e^{n\sigma  + \Upsilon } \hat \epsilon \,\,\,\,{\rm where\,\,} \hat \epsilon\,\, {\rm{is\,\, a\,\, constant\,\, spinor,}}\nonumber\\
    n &=& \frac{1}{2}\left[ { \pm 1 + i\alpha \left( { \pm 1 - \frac{1}{2}} \right)} \right]\,\,\,\, {\rm and}\nonumber\\
    \left( {\partial _\mu  \Upsilon } \right) &=& \frac{1}{4}\left[ {Y_\mu \pm \left( {\partial _\nu  \sigma } \right)\varepsilon _\mu ^{\,\,\,\,\,\nu } } \right].
\eea

If we now solve the Laplace equation $\Delta H=0$ to find $H\left( r\right) = 1 + \frac{q}{{r }}$, where $q \in \mathbb{R}$ and $r$ is the usual radial coordinate in 3-d space, then:

\bea
    ds^2  &=&  - dt^2 + dx^2  + \left( {1 + \frac{q}{{r }}} \right)\left( {dr^2  + r^2 d\Omega ^2_2} \right)     \nonumber\\
    \sigma \left( r \right) &=& \ln \left( {1 + \frac{q}{{r }}} \right), \quad\quad\quad
    a = a_\infty   \pm \frac{{7q}}{{2\sqrt 7 \left( {r  + q} \right)}}\label{metricanddilaton}\\
    dz^i  &=& - q f^i\frac{{dr}}{{r^2 }}\,\,\,\,\,\,{\rm such}\,\,{\rm that}\,\,\,\,\,\,df^i  - q\Gamma _{jk}^i f^j f^k\frac{dr}{r^2} =0\,\,\,\,\,\,{\rm and}\,\,\,\,\,\,G_{i\bar j} f^i f^{\bar j}  = \frac{{r^2 }}{{\left( {r  + q} \right)^2 }}    \label{Affine}\\
    \left| \Xi  \right\rangle  &=&   \sqrt {\frac{r}{{7\left( {r + q} \right)}}} \mathfrak{Re}\left[ {\left( {1 \pm i\sqrt 7 } \right)\left| V \right\rangle } \right] + 2\sqrt {\frac{{r + q}}{{7r}}} \mathfrak{Re}\left[ {f^i \left| {U_i } \right\rangle } \right]\nonumber\\
    \left| {d\Xi } \right\rangle  &=& \pm q\mathfrak{Re} \left[ {\frac{{\left( { \pm \sqrt 7  - i} \right)}}{{\sqrt {r\left( {r + q} \right)^3 } }}\left| V \right\rangle  + \frac{{2i}}{{\sqrt {r^3 \left( {r + q} \right)}  }}f^i \left| {U_i } \right\rangle } \right]dr,\label{FullSolution}
\eea
where $d\Omega ^2_2$ is the unit $S^2$ metric. Equations (\ref{metricanddilaton}), (\ref{Affine}) and (\ref{FullSolution}) represent a full symplectic solution, but only a partial spacetime one. The entire construction is based on the choice that the dilaton and the universal axion are independent of the moduli, and that the entire moduli dependence is carried exclusively by the axions, while the moduli themselves are dependent on an unknown symplectic scalar $f^i$. The conditions we found on these functions can be understood in the following way: Based on the first equation of (\ref{Affine}), one can define $\lambda  = {q \mathord{\left/ {\vphantom {q r}} \right. \kern-\nulldelimiterspace} r}$, such that
\be
    f^i  = \frac{{dz^i }}{{d\lambda }},
\ee
\emph{i.e.} the scalar functions $f^i$ represent ``velocities'' on the space of complex structure moduli $\M_C$ with $\lambda$ acting as an affine parameter. The normalization of these velocities is given by the last of equations (\ref{Affine}), which can be rewritten as follows:
\be
    G_{i\bar j} \left( {\frac{{dz^i }}{{d\lambda }}} \right)\left( {\frac{{dz^{\bar j} }}{{d\lambda }}} \right) = \frac{1}{{\left( {1 + \lambda } \right)^2 }}.
\ee

In this language, one can think of the middle equation of (\ref{Affine}) simply as the geodesic equation on $\M_C$, describing the flow of the spacetime solution on the space of complex structure moduli of the underlying submanifold:
\be
    \frac{{d^2 z^i }}{{d\lambda ^2 }} + \Gamma _{jk}^i \left( {\frac{{dz^j }}{{d\lambda }}} \right)\left( {\frac{{dz^k }}{{d\lambda }}} \right) = 0.
\ee

This interpretation clearly provides grounds for future study. While a complete solution requires full knowledge of at least a metric on $\M_C$, study of these equations could further specify this and similar solutions (such as \cite{Emam:2012zw}).

Finally, we note that the quantity $q$ is a coupling constant relating the behavior of the fields to each other and to gravity. Since the metric is asymptotically flat; the ADM mass of the brane is easily calculable and is clearly proportional to $q$. Since the value of $q$ can be either positive or negative, we note the following: For positive values of $q$ the solution is entirely smooth between the central singularity and infinity. While for the case of negative $q$, a curvature singularity exists at $r={\left| q \right|}$. As such the negative $q$ result has two singularities, one at $r=0$ and the other constituting an $S^2$ surface with radius $r={\left| q \right|}$. In both cases the singularities are naked; no horizons exist.

\section{Conclusion}

This work is a continuation of the results presented in our previous paper \cite{Emam:2012zw}, whose primary objective was to apply the methods developed in \cite{Emam:2009xj} and construct $D=5$ hypermultiplet fields in a specific spacetime background, simply by exploiting the symplectic symmetry of the theory and finding solutions that are based on symplectic invariants and vectors. In so doing, we have also shown that only two (Poincar\'{e})$_{p+1} \times SO\left(4-p\right)$ backgrounds are allowed (within the symmetries assumed). Focusing here on the second of these possibilities, we constructed a BPS one-brane coupled to the hypermultiplet fields of $\N=2$ supergravity. The metric and fields are well-behaved in the far field region and are dependent on the ADM mass density of the brane. We found explicit expressions for the metric, dilaton and the universal axion. On the other hand the axions are dependent on spacetime-unspecified symplectic basis vectors and the moduli are proportional to an unknown set of functions $f^i$ describing the flow of the solution along the space of complex structure moduli of the underlying CY submanifold. The geodesic equation of such a flow was derived, as well as the normalization condition of $f^i$. What we have then is a complete symplectic solution, but a partial spacetime one. Clearly, a full solution hinges on the values of $f^i$, \emph{i.e.} on solving the aforementioned constraints. Also, further study of these functions may provide clues to the space of complex structure moduli of the CY submanifold. One other possible direction of future research is studying the M-theoretic interpretation of these results, \emph{\emph{i.e.}} lifting the result to eleven dimensions.

\pagebreak

\end{document}